\documentstyle[11pt]{article}
\newcommand{\beq}{\begin{equation}}
\newcommand{\enq}{\end{equation}}
\newcommand{\lan}{\langle}
\newcommand{\ran}{\rangle}
\begin{document}

\begin{titlepage}
\begin{flushright}
		Preprint KUL-TF-93/20 \\
		hepth@xxx/9305149\\
		26.05.1993 \\
\end{flushright}
\vfill
\begin{center}
{\large\bf Non-abelian Harmonic Oscillators and Chiral Theories} \\
\vskip 27.mm
{\bf Z. Hasiewicz\footnote{Onderzoeker I.I.K.W. Belgium,
on leave from IFT University of Wroclaw, Poland}
, P. Siemion\footnote{On leave from IFT University
of Wroclaw, Poland. \\ Partially supported by KBN-grant 2 00 95 91 0 }}\\
\vskip 1cm
Instituut voor Theoretische Fysica
	\\Katholieke Universiteit Leuven
	\\Celestijnenlaan 200D
	\\B-3001 Leuven, Belgium\\[0.3cm]
\end{center}
\vfill

%% \vfill
%%\title{NON-ABELIAN HARMONIC OSCILLATORS AND CHIRAL THEORIES }
%%\author{ Z.Hasiewicz , P. Siemion

%%\maketitle

\begin{abstract}
We  show  that  a  large  class  of   physical   theories
which has been under intensive investigation recently, share the
same geometric features in their Hamiltonian formulation. These dynamical
systems range from harmonic oscillations to WZW-like models and to the KdV
dynamics on $Diff_oS^1$. To the same class belong also the Hamiltonian
systems on groups of maps. \\
 The common feature of these models are the 'chiral' equations of motion
allowing for so-called  chiral decomposition of the phase space.
\end{abstract}
\vskip 27.mm
\thispagestyle{empty}

\end{titlepage}

\section{Introduction}
It is common impression that many of  geometrical  models
in Field Theory share some common and universal properties,
despite the technical complexity of descriptions of the models.
For each of these models a specific 'machinery' has been developed,
making the common features less visible. Maybe the most prominent example
of  this class is the Wess-Zumino-Witten  Field   Theory   \cite{15}.
Both  classical Hamiltonian formulation \cite{11} and corresponding
quantum  theory \cite{5}  are being viewed on the ground of the results
which are known or are expected to be obtained. This indicates that
in  spite  of  the  great development done  recently \cite{3}
\cite{6}\cite{4}, some fundamental background is missing.
\par
It is our hope to shed some new light on those questions by showing
a large family of the models, ranging from harmonic oscillations and
free motions to the dynamical systems on the groups  of maps or groups
of diffeomorphisms, and making the common features evident.
In fact all those models turn out to be straightforward generalizations
of harmonic oscillations and free motion. \\
Better understanding of the geometrical nature of those models may be very
helpful in quantization, as for instance it allows one to use  the data of the
representation theory in a more conscious (and efficient) way.
\par
The starting point for definition of this class of models is the group
structure on the corresponding configuration space. The next ingredient
is the non-canonical lifting of the left and right actions of the group
on itself to the cotangent bundle (the phase space). We equip the phase space
with a symplectic form which is invariant under the above (lifted) actions
and allows for their Hamiltonian realisation (by momentum mappings).
\par
The Hamiltonian defining the dynamics is just a quadratic function of these
momentum mappings, and this guarantees that the equations
of motion are 'chiral'.

\par
The paper is organized as  follows:  the  first  section
contains a group-theoretical approach to the case of  standard
harmonic oscillators and free motions. Its power  consists  on
its staightforward generalizations to the class  of
models describing motions on group manifolds; the Hamiltonian
description of these theories is contained in next sections.
\par
The third section is devoted to presentation of some more important
and more complicated examples (like WZW model and the dynamical
model on $Diff_oS^1$ - which leads to chiral KdV equations for
the momentum mappings).
\par
In the summary we present briefly the geometrical nature of the chiral
splitting.

\section{The harmonic oscillation and free motions}
In accordance with our promise made  in  the  introduction  we
shall briefly reformulate the theory of harmonic oscillators.
\par
Let us consider  the $n$-dimensional  real  space  as  the
configuration space of the model. This space has  a  structure
of a $n$-dimensional abelian (additive) group.
\par
The left and right actions of this group on itself can be
lifted to the action on the phase space, but this  lifting  is
not unique. We shall consider the following actions:
\par
\beq
\Phi ^{l}_{a}(x,p) := (x+a\,,\,p + La)
\label{1}
\enq
\beq
\Phi ^{r}_{a}(x,p) := (x-a\,,\, p + Ra)\label{2}
\enq
\par
 where $(x,p)$ describes the position  and  momentum
and $a$ is a (vector) parameter of translation. $L$ and $R$
are linear operators from the group to its dual. The above actions are
nothing  else  but affine extensions of left and right translations by
means  of the group cocycles $\theta ^{l(r)}(a) := L(R)$a.
\par
 By hand (at the moment) we shall equip the  phase  space  with
the following (non-canonical) symplectic structure:
\par
\beq
\Omega  := \lan dp , \wedge dx \ran + {1\over 2}\lan
(R^{A} - L^{A})dx , \wedge dx \ran
\label{3}
\enq
 where $\langle$ , $\ran$ stands for the pairing and $R^{A}$ and
$L^{A}$ stand for antisymmetric parts of $R$ and $L$ with respect to the
pairing. The symbol $dx$ should be understood as a vector-valued one-form
(n-bein).
\par
 It  is  easy  to  check  that  canonical  variables  have  the
following Poisson brackets:
\par
\beq
\{ p_i , x^j \} = \delta _i^j
\label{4}
\enq
\beq
\{ p_i , p_j \} = (R^{A} - L^{A})_{i,j}
\enq
\beq
\{ x^{i} , x^{j} \} = 0 .
\enq
 It is also easy to check that the  actions  (\ref{1})  and  (\ref{2})  are
Hamiltonian and admit the following momentum mappings \cite{7}:
\par
\beq
J^{r}(x,p) = p + ( R^{S} - L^{A} )x
\label{jra}
\enq
\beq
J^{l}(x,p) = -p + ( L^{S} - R^{A})x,
\label{jla}
\enq
 where the letters with superscripts $S$  denote  the  symmetric
parts of corresponding operators. To avoid  confusion  of  the
momentum mappings with the canonical momenta we shall call the
former (abusing the terminology slightly) the chiral momenta.
\par
 Using (\ref{4}) one can find that:
\par
\beq
\{ J^{r}_{i} , J^{r}_{j} \} = (R^{A} + L^{A})_{ij}
\label{6}
\enq
\beq
\{ J^{l}_{i} , J^{l}_{j} \} = - (R^{A} + L^{A})_{ij}
\label{7}
\enq
\beq
\{ J^{r}_{i} , J^{l}_{j} \} = - (R^{S} + L^{S})_{ij} .\label{8}
\enq
 Let us notice that in spite of commutativity of the Lie
algebra of translations its hamiltonian realization gets
centrally extended. Such a realization is called weakly
Hamiltonian sometimes \cite{10}. In the next section we shall assume
that the symmetric parts of the cocycles vanish, thus making
left and right currents decoupled.
\par
The hamiltonian of the theory is simply the sum of the
sqares of chiral momenta:
\beq
{\cal H} := {1 \over 4} [ ( J^{l} , J^{l} ) + ( J^{r} , J^{r} ) ]
	\label{9}
\enq
where ( , ) stands for the Euclidean scalar product.
A straightforward calculation shows that
\par
\beq
{d^{2}\over dt^{2}} x = - {1\over 4} (R^{t}+L^{t})(R+L)x
\label{10}
\enq
 where $t$ stands for transposition with respect to ( , ). Let us notice
that the matrix on the r.h.s. of (\ref{10}) is never positive (as we assumed
that  ( , ) is Euclidean). This means
that the second order equation describes either oscillating
(positive eigenvectors ) or free (null vectors) motions.
\par
It is easy to verify that the equations of motion for $J^{l}$
and $J^{r}$ generated by (\ref{9}) are:
\par
\beq{d\over dt} J^{l} = {1 \over 2} (L+R) \hat J^{l}
\label{11}
\enq      %%%
\beq
{d\over dt} J^{r} =- {1 \over 2} (L+R) \hat J^{r}
 \label{12}
\enq
where $\hat J^{l,r}$ is the vector dual to $J^{l,r}$ via ( , ).
As we shall see later, they are in precise analogy with the
equations of motion for chiral currents in  the WZW model \cite{6} .
This is the very reason why we call $J^r$, $J^l$ the chiral momenta.

\section{The general case}
 We can now proceed to a much more general case: let us assume
that the configuration space is an arbitrary group manifold $G$
 (in particular infinite-dimensional). Moreover, we shall
assume that the objects we are intended to consider do exist (
even in the infinite-dimensional case). This is automatically
satisfied in the cases of groups of maps from
compact manifolds (with the loop groups as a special case
corresponding to the WZW theory).
\par
In order to parametrize the cotangent bundle we shall
trivialize the bundle by means of the left action of the group.

\par
The points of the phase space $(T^*G)$ are thus described by
pairs $(g,p)$ where $g \in  G$ and $p \in  {\cal G}^*$ . The left and right
action of the group on itself can be lifted to the action on the
phase space in the following way:
\par
\beq
\Phi ^{l}_{g_{o}}(g,p) := ( g_{o}g , p + Ad^{*-1}_{g_{o}g}\theta ^{l}(g_{o}) )
\label{13}
\enq
\beq
\Phi ^{r}_{g_{o}}(g,p) := ( gg^{-1}_{o}, Ad^*_{g_{o}}p + \theta ^{r}(g_{o}) )
\label{14}
\enq
 where $\theta ^{l}$ and $\theta ^{r}$ are arbitrary ${\cal G}^*$-valued
group cocycles, i.e. they do satisfy:
\beq
\theta (g_{1}g_{2}) = Ad^*_{g_{1}}\theta (g_{2}) + \theta (g_{2}) ,
\label{15}
\enq
and $Ad^*$ is the coadjoint action.
\par
 The formulas (\ref{13}) and (\ref{14}) are in exact correspondence to
those of (\ref{1}),(\ref{2}). They do not look symmetrically because we
are using the left trivialization. If we used the right
trivialization instead, the nonlocality would appear in the
expression for $\Phi ^{r}$.
\par
The canonical symplectic structure of the cotangent
bundle is given by the differential of the Liouville form:
\par
\beq
\Omega  := d\alpha  ;\quad \hbox{  where }\alpha  :=
\lan p , g^{-1}dg \ran
\label{16}
\enq
 and obviously the pairing is evaluated in ${\cal G}$ - the space of
values of the canonical left-invariant form $g^{-1}dg$.
\par
 It is not difficult to check that neither the Liuville form
nor its differential are invariant under (\ref{13}),(\ref{14}).
This means that the actions (\ref{13}, \ref{14}) cannot be realized in
a Hamiltonian way.
\par
 In order to obtain an invariant symplectic form one has to add
to $\Omega $ an additional term:
\par
\beq
\Omega _{\hbox{inv}} := \Omega  - \lan \Sigma ^{l}(dg\,g^{-1}) , \wedge
dg\,g^{-1} \ran + \lan \Sigma ^{r}(g^{-1}dg),\wedge g^{-1}dg \ran
\label{17}
\enq
 By $\Sigma ^{l,r}$ we understand the derivative of the cocycles
$\theta ^{l,r}$ at the group unity. By definition they are linear
operators on ${\cal G}$.
At this point we make an assumption that these operators are
$\lan$ , $\ran$ -antisymmetric. In the literature \cite{13} the
cocycles with antisymmetric derivatives are called symplectic.
\par
 Let us notice that we are free to add to (\ref{17}) an arbitrary
closed and both left- and right-invariant 2-form on $G$. On  a
semi-simple  group however (contrary to the abelian case
considered in the section 2) the only one such form is 0
\cite{Bour}.
\par
The actions (\ref{13}),(\ref{14}) are Hamiltonian with respect to
(\ref{17}) and admit the following (weak) momentum mappings:
\par
\beq
J^{l}(g,p) := -Ad^*_{g}p - \theta ^{r}(g)
\label{18}
\enq
\beq
J^{r}(g,p) := p - Ad^*_{g^{-1}}\theta ^{l}(g) .
\label{19}
\enq
\par
The Poisson algebra of (\ref{18}) and (\ref{19}) with respect to
$\Omega_{inv}$ has the following form:
\par
\beq
\{J^{l}(X),J^{l}(Y)\} = J^{l}([X,Y]) + C^{+}(X,Y)
\label{20}
\enq
\beq\{J^{r}(X),J^{r}(Y)\} = J^{r}([X,Y]) - C^{+}(X,Y)\label{21}
\enq
\beq
\{J^{l}(X),J^{r}(Y)\} = 0
\label{22}
\enq
 where
\beq
C^{\pm }(X,Y) := C^{l}(X,Y) \pm  C^{r}(X,Y) ,
\label{23}\enq
\beq
C^{l,r}(X,Y) := \lan \Sigma ^{l,r}(X) , Y \ran
\label{24}
\enq
 and $\Sigma ^{l,r}$ is a derivative of a respectve cocycle as in (\ref{17}).
\par
Let us notice here   that   only   the   sum   of   the
cocycles matters  for  the  central  extensions  of   the   Lie
algebras of left and right translations. In the  standard  WZW
theory one assumes from the very beginning that $C^{-} = 0$ .
\par
In order to induce the dynamics let us introduce the following quadratic
Hamiltonian:
\beq
{\cal H} := {1\over 4} \big ( K(J^{l}, J^{l}) + K(J^{r}, J^{r}) \big)
	\label{27}
\enq
where $K$ is some quadratic form on ${\cal G}^*$.
If we assume that $K$ is $Ad^*$-invariant, then (\ref{27}) together with
(\ref{17}) give the following equations of motion:
\par
\beq
{d\over dt} J^{l}= {1\over 2} \Sigma ^+ (\hat J^{l})
\label{25}
\enq
\beq
{d\over dt} J^{r} = - {1\over 2} \hat \Sigma ^+ (\hat J^{r}) .
\label{26}
\enq
where
\beq
	\Sigma ^\pm  = \Sigma ^l \pm \Sigma ^r
\enq
and $\hat J^{l,r}$ is the $K$ - dual of $J^{l,r}$ .
\par
 These equations we shall call chiral, as  in  the  case  of $G$
being a loop group they are precisely the equations of  motion
for the chiral currents in WZW theory (see below).
\par

The equations of motion for a group point are simple:
\par
\beq
{d\over dt}\  g = {1\over 2}\ (g \hat J^{r} - \hat J^{l}g)
	\label{28}
\enq
whereas for the canonical momenta one has:
\beq
{d \over dt} p =-{1 \over 2} \Big( ad^*_{\hat J^r} p + \Sigma ^r (\hat J^r)
		+ Ad^*_{g^{-1}} \Sigma ^l (\hat J^l) \Big ) =
\enq
\beq
	=       - {1 \over 2} \Big ( \Sigma ^r (\theta ^l (g^{-1})) +
		\Sigma ^l (\theta ^r (g^{-1})) - \Sigma ^- (p) +
		[\theta ^l (g^{-1}) , \theta ^r (g^{-1}) - 2p] \Big )
\label{28b}
\enq
The equations (\ref{28},\ref{28b}) seem to be difficult to solve
explicitely and in any case are much more complicated than
(\ref{25},\ref{26}). On the other hand one can observe that the
Hamiltonian is in fact a collective one in the sense of \cite{7},
i.e. it is a pull-back of a function on ${\cal G}^* \times {\cal G}^*$
by the chiral momenta (\ref{18},\ref{19}). This fact has profound
geometric consequences and we shall shed some light upon it in one
of the following sections

\section{Examples}
To ilustrate the theory let us describe briefly some
examples. \\
\noindent
1) The case of finite-dimensional Lie group.
\par
As the first instance one  can  consider  a  prticle
moving on a semi-simple Lie group manifold. In this  case  all
cocycles are trivial \cite{Bour} , i.e. of the form:
\par
\beq
\theta ^{l,r}(g) = Ad^*_g \mu ^{l,r}-\mu ^{l,r} ;\quad
\mu ^{l,r}\in {\cal G}^*
\label{29}
\enq
 and then
\beq
	\Sigma ^{l,r} =  ad^*_{g^{-1}dg} \mu ^{l,r}
	\label{30}
\enq
 The invariant symplectic form is simply:
\par
\beq
\Omega _{\hbox{inv}} = d \alpha  + \lan \mu ^{l},dg\,g^{-1}\wedge
dg\,g^{-1}\ran-
\lan \mu ^{r}, g^{-1}dg \wedge g^{-1}dg \ran
\label{31}
\enq
\par
The equations of motion generated by the  Hamiltonian  of
(\ref{27}) are analogous to the equations of a particle moving in  a
magnetic-like field of strenght $\mu ^{+}$.
The equations of motion (\ref{28},\ref{28b}) can be solved explicitely
on any semi-simple Lie group \cite{forth},
and by 'explicitely' we do not mean the time-ordered integrals, but rather
analytical expressions depending on time and initial data (position and
momentum). The trick in solving the equatins consists in proper use
of chiral factorization (see below).
In particular  for $\mu ^{+} = 0$ they describe the
free motion (along the big circles for the compact $G$) as in \cite{9} .
\par
\medskip
\noindent
2) Loop groups and WZW.
\par
Another example is given by the configuration space being the loop
group ${\cal L}G$. In this case, since the group is infinite-dimensional,
we should from the very
begining extract from ${\cal LG}^*$ the smooth part \cite{12} by
identifying it with ${\cal LG}$ via
the non-degenerate form on the  Lie  algebra ${\cal LG}$  ,
defined as follows:
\beq
K(X,Y) = {1\over 2\pi }\int^{2 \pi}_{0} {\cal K}(X(\sigma ),Y(\sigma ))
d\sigma
\label{32}\enq
 where ${\cal K}$ is an Ad-invariant form on ${\cal G}$.
In  this  case  we  will take non-trivial cocycles on ${\cal L}G$:
\beq
\theta ^{l,r}(g) = k^{l,r}({d\over d\sigma }g)g^{-1} ;
k^{l,r}\in {\bf R}
\label{33}
\enq
 and then
\beq
\Sigma ^{l,r} = k^{l,r} {d\over d\sigma }
\label{34}
\enq
The chiral currents  do  satisfy  the  following  Poisson
comutation relations:
\par
\beq
\{ J^{l}(X),J^{l}(Y) \} =
J^{l}([X,Y])-(k^{l}+
k^{r})\int^{2\pi }_{0} {\cal K}(X(\sigma ),
{d\over d\sigma }Y(\sigma )) d\sigma
\label{35}
\enq
\beq
\{ J^{r}(X),J^{r}(Y) \} =
J^{r}([X,Y])+(k^{l}+k^{r})
\int^{2\pi }_{0}{\cal K}(X(\sigma ),
{d\over d\sigma }Y(\sigma )) d\sigma
\label{36}
\enq
 which  are  easily  recognized  as  the   affine   algebras
with underlying Lie algebra ${\cal G}$.
\par
By introducing the chiral derivatives:
\beq
\partial _{\pm } := {\partial \over \partial t}
\pm  {1\over 2}(k^{l}+k^{r}) {\partial \over \partial \sigma }
\label{37}
\enq
 we can write the chiral equations of motion as:
\beq
\partial _{+} J^{l} = 0 = \partial _{-} J^{r} .
\label{38}
\enq
 The chiral derivatives (\ref{37}) one can get in a standard form  by
appropriate  redefinition  of  time  variable  or   equivalent
rescaling of the Hamiltonian.
\par
Similarly one could introduce the chiral dynamics on an arbitrary group of
maps ${\cal M}G$, but this time the cocycle would not be invariant
under the group of diffeomorphisms of $M$. In fact all the cocycles are
defined by the 1-cycles on $M$ in this case. Such models would be of
considerable physical importance as they could describe the motions in the
presence of monopole-like singularities in $M$.
\par
\medskip
\noindent
3) $Diff_oS^1$ and KdV.
\par
As another example let us consider $G = Diff_o S^1$ - the group of
orientation-preserving diffeomorphisms of $S^1$. The tangent space at
identity is then the space of vector fields on $S^1$ \cite{Kir}:
\beq
	V = T_{id}G \ni \xi = \xi (\sigma) {d \over d\sigma} ;
	\quad \hbox{where $\sigma$ is a coordinate on }S^1
\enq
and the space of smooth moments can be identified with the space of
quadratic differentials
\beq
	V^* = T_{id}^*G_{smooth} \ni p = p(\sigma) d\sigma \otimes d\sigma
\enq
The pairing is defined as contraction and then integrating over $S^1$:
\beq
	\lan p , \xi \ran = \int_0^{2\pi} p(\sigma) \xi (\sigma)
	d\sigma
\enq
and is clearly invariant under reparametrizations of $S^1$.
\par
	The main new feature is however the lack of a both left- and
right-invariant form on $V$. This means that one has to be very
careful about whether he is working in $V$ or in $V^*$, contrary to the
case of a Lie group, where the identification between ${\cal G}$ and
${\cal G}^*$ defined by $K$ allows for some carelessness in this respect.
\par
Out of many non-invariant quadratic forms on $V^*$ let us choose the simplest
one:
\beq
	K(p,q) = \int_0^{2\pi} p(\sigma) q(\sigma) d\sigma
	;\quad \forall p,q \in V^*
	\label{kk}
\enq
$K$ can be equivalently considered to be a mapping
\beq
	K: V^* \ni p \mapsto \hat p = p(\sigma)
	{\partial \over \partial\sigma}\in V .
\enq
The Hamiltonian defined as in (\ref{27}) is not $Ad$- invariant, as it is
defined by a non-invariant form (\ref{kk}).
\par
Now we have to choose $\theta ^l$ and $\theta ^r$.
The non-trivial cocycle is given by \cite{Bott}:
\beq
	\theta (\phi) = Ad^*_\phi S(\phi ) ; \quad \phi \in Diff_o S^1,
\enq
where $S$ is the Schwarzian:
\beq
	S(\phi) = {{\phi '\phi ''' - 3/2 (\phi '')^2} \over {(\phi ')^2} }
	d\sigma \otimes d\sigma
\enq
and the primes denote differentiation. \\
As in general this is the sum $\theta ^r + \theta^l$ that is relevant for
the dynamics, let us take $\theta ^{r+l} \equiv \theta$. \\
The derivative of $\theta$ is
\beq
	\Sigma (\eta) = \eta(\sigma)''' d\sigma \otimes d\sigma \quad;
	\quad \forall \eta \in V
\enq
and the 2-form $C$ is easily seen to be the Gel'fand - Fuks cocycle
\cite{Gel}:
\beq
	C(\xi,\eta) := \lan \Sigma (\xi) , \eta \ran = \int_{S^1}
	\xi(\sigma)'''\eta(\sigma) d\sigma = - \int_{S^1} \xi(\sigma)'
	d \eta(\sigma)'
\enq
Now let us calculate the coadjoint action of $V$ on $V^*$:
\beq
	\lan ad^*_{\xi}p , \eta  \ran := \int_{S^1} p [\eta,\xi] =
	\int_{S^1}  \big ( 2\xi(\sigma)'p(\sigma)+ \xi(\sigma)p(\sigma)'
	\big )  \eta(\sigma) d\sigma
\enq
and as this is satisfied for any $\eta \in V$, we have
\beq
	ad^*_\xi p = \big ( 2\xi(\sigma)'p(\sigma)+ \xi(\sigma)p(\sigma)'
	\big )  d\sigma \otimes d\sigma .
\enq
In particular for any $p \in V^*$
\beq
	ad^*_{\hat p}p = 3p(\sigma)p(\sigma)'d\sigma\otimes
	d\sigma \neq 0 \hbox{ (!!!)} .
\enq
This is a consequence of $K$ being non-invariant.\\
The equation of motion for the right chiral momentum $J^r$ is again
\beq
	{d \over dt} J^r = -{1 \over 2} \Sigma(\hat J^r)
		-{1 \over 2}  ad_{\hat J^r}^*J^r
\enq
but this time the last term does not vanish (compare (\ref{26})).
In terms of functions this reads:
\beq
	\dot J^r  + {1 \over 2} (3J^r(J^r) ' + (J^r)''') = 0
	\label{kdvr}
\enq
which is nothing else but the equation of Korteweg - de Vries. \\
Similarly for the left momentum we get:
\beq
		{d \over dt} J^l = {1 \over 2} \Sigma(\hat J^l)
		+{1 \over 2}  ad_{\hat J^l}^*J^l
\enq
or in the coordinates
\beq
	\dot J^l - {1 \over 2} (3 J^l(J^l)' +(J^l)''') = 0
	\label{kdvl}
\enq
Comparing (\ref{kdvr}) and (\ref{kdvl}) with the case of loops (\ref{38})
we see that the chiral equations has been replaced by higher equations
of KdV hierarchy. The equations are not linear any longer, but the
nonlinearity is in a sense 'minimal', as the equations of KdV are completely
solvable. The nonlinear terms have their origin precisely in the
non-invariance of the Hamiltonian.

\section{Concluding remarks}
In order to summarize the presented formulation and the examples we shall
make some remarks about the underlying geometry.
\par
Let us go back to the abelian case considered in the section 2.
The crucial object there is the matrix $L+R$. This matrix is in general
noninvertible (as we have seen in section 3. it is genericly antisymmetric)
and threfore the cofiguration space splits into its null
subspace and the subspace $M_{invert}$ on which $L+R$ is invertible.
{}From the equations (\ref{jra}),(\ref{jla}) one immediately sees that
knowing $J^r$ and $J^l$ it is possible to recover the information
about position up to an arbitrary translation along the null space. \\
On the other hand from the time evolution of chiral momenta (\ref{11})
(\ref{12}) it is clear that it is precisely the restriction to
$T^*M_{invert}$ where we observe oscillations, while the motion in the
complementary space is free. Therefore we can effectively describe
the phase space by the chiral momenta plus initial position and velocity
in the null sector. In three dimensions it has a nice interpretation as
decomposition into left- and right circular polarisation.
\par
Let us now discuss the general case. \\
As we have already noticed, the Hamiltonian is a pull-back of a function
on ${\cal G}^* \times {\cal G}^*$
by the chiral momenta (\ref{18},\ref{19}).
This means that one can try
to describe the motion in terms of ${\cal G}^*$ - valued momenta as their
equations of motion are much simpler. Thus one should consider the image
of $T^*G$ in ${\cal G}^* \times {\cal G}^*$ under the map
\beq
	J := J^r\times -J^l.
	\label{J}
\enq
{}From the definitions (\ref{18},\ref{19}) it follows that the chiral
momenta are not independent, or to be more precise:
\beq
	-J^l(g,p) = Ad^*_g(J^r(g,p)) + \theta ^+ (g)
	\label{affin}
\enq
i.e. they are on the same orbit of the affine action defined by the cocycle
$\theta ^+$:
\beq
	{\cal A}^+ (g) = Ad^*_g + \theta ^+ (g)
	\label{A}
\enq
Therefore the image of $T^*G$ under (\ref{J}) is precisely the fibered
product defined by projection $\pi$ on
the space of affine orbits:
\beq
	J(T^*G) = {\cal G}^* \times_{\pi} {\cal G}^*
\enq
where $\pi$ assigns to each element $\xi \in {\cal G}^*$ its conjugacy
class with respect to the action ${\cal A}^+$. It is clear that the
point in a fibred product contains less information about the system
than a point in $T^*G$. In order to see what we have lost let us look
at the fibre of $J$ over the point $(\xi, \xi ') \in {\cal G}^* \times_{\pi}
{\cal G}^*$. The fibre is clearly isomorphic to the affine stabilizer
subgroup of $\xi$ (or any other point conjugated with $\xi$ by ${\cal A}^+$).
\par
Let ${\cal W}$ be the set of conjugacy classes (affine orbits). Defining
the projection $\pi_J := \pi \circ J$ of $T^*G$ on ${\cal W}$ one can see that
for any  point  $r \in {\cal W}$
\beq
	\pi_J^{-1} (r) \cong H \times G/H \times G/H
	\label{kanap}
\enq
where $H$ is the affine stabilizer of some $\eta$ in class $r$.
Obviously $H$ as a subgroup depends on the choice of $\eta$ ($r$ fixes it up
to an isomorphism only). One can thus immediatly see the possible
obstructions for the fibres to fit together into the structure of
a smooth bundle over ${\cal W}$. However, the situation is not hopeless
because in many interesting cases the mapping $\pi$ defines a trivial
fibration over the open subset ${\cal W}_o$ of ${\cal W}$. The inverse image
$\pi_J^{-1}({\cal W}_o) =: (T^*G)_o$ is called the set of regular points.
\par
In the case of compact Lie groups or loop group the space ${\cal W}_o$ can
be identified with the interior of the Weyl Chamber in $t^*$ , where $t$ is
the Lie algebra of some maximal torus of $G$. Then the layers of (\ref{kanap})
fit together to give the principal bundle structure:
\beq
H \hookrightarrow (T^*G)_o \buildrel J \over\rightarrow ({\cal G}^*
\times_{\pi} {\cal G}^* \simeq G/H \times G/H) \buildrel
\pi\over\rightarrow
{\cal W}_o
\enq
This is not the end of the story, because the above structure factorizes.
Let us consider two copies $P_{l,r}$ of the manifold ${\cal W}_o \times G$.
endowed with the following symplectic structures:
\beq
	\Omega_l = d \lan r_l, g^{-1}_l dg_l \ran + {1 \over 2} \lan
	\Sigma ^+ (g^{-1}_l dg_l),g^{-1}_l dg_l \ran
\enq
\beq
	\Omega_r =  d \lan r_r, dg_r g^{-1}_r \ran + {1 \over 2} \lan
	\Sigma ^+ (dg_r g^{-1}_r),dg_r g^{-1}_r \ran
\enq
and let $P = P_l \times P_r$ be a symplectic manifold with the form
$\hat \Omega = \Omega_l - \Omega_r$.
\par \noindent
Then
\beq
	(T^*G_o , \Omega) \simeq (P,\hat \Omega) / (r_l - r_r = 0)
\enq
where $/$ denotes the symplectic reduction by the set of (first class)
constraints. The above statement we call the chiral splitting because
$P_{l,r}$ describe the dynamics of chiral momenta. The elements $g_{l,r}$
correspond to vertex operators of WZW.
\par
In the particular case of ${\cal L}G$ the form $\Omega_l$ has the following
structure:
\beq
\matrix{ \Omega_l = d \lan r_l, g^{-1}_l dg_l \ran + {1 \over 2} (k^l+k^r)
	\lan (u^{-1} du)', u^{-1} du \ran + \cr \cr
	+ \, d \lan r_l , Ad_{g_l}^{-1} u^{-1}du \ran }
	\label{sect}
\enq
where $g_l$ is a constant loop (the zero mode) and $u$ is an element of
the group of loops based at the unity. The second term in $\Omega_l$
describes the canonical symplectic structure on the manifold of based loops.
It is known that this manifold admits a complex structure which is compatible
with the symplectic one \cite{12} and therefore it defines the Kaehler
structure. This property is very important for quantization.
{}From (\ref{sect}) one can see that kinematically the chiral sector splits
into so-called zero-modes and higher, oscillating modes, described by based
loops. The zero-mode dynamics is nothing else but a chiral part of free
motion of a 'point particle' on a group manifold \cite{9}.
The third term of (\ref{sect}) couples the zero-modes to the oscillating
ones. Its presence forces the so-called screening, i.e. it sets the upper
limit for values of the variable $r_l$ in terms of $k_l+k_r$. It is only
with this restriction that the symplectic form is non-degenerate.
Because $r_l$ labels the representations of the current algebra the
above condition restricts the 'spin' content of the theory.
\par
It would be interesting to perform a similar analysis for the case of
$Diff_o S^1$. In this case the structure of the orbit space is a little bit
more complicated \cite{Kir} \cite{Witten}, but still controlable. This
gives one some hope that the canonical quantization of the corresponding
theory  can be performed. We shall return to this issue in the future paper,
as well as to the question of the canonical geometry of groups of maps.

\section{Acknowledgements}

The authors would like to thank the all members of the Institute of
Theoretical Physics of K.U.Leuven for their hospitality. One of the authors
(P.S.) would like to thank Alberto Verjovski for a discussion.

\end{document}